\documentstyle[11pt,aaspp4,flushrt]{article}

\newcommand{\apg}{\gtrsim}
\newcommand{\apl}{\lesssim}

\newcommand{\lya}{\mbox{${\rm Ly}\alpha$}}
\newcommand{\lyb}{\mbox{${\rm Ly}\beta$}}

\begin{document}

\title{PHOTOMETRIC REDSHIFTS OF GALAXIES IN THE HUBBLE DEEP FIELD}

\author{KENNETH M. LANZETTA, ALBERTO FERN\'{A}NDEZ-SOTO\altaffilmark{1}, and
AMOS YAHIL
\\ Department of Physics and Astronomy, State University of New York at Stony
Brook \\
Stony Brook, NY 11794--3800, U.S.A.}

\altaffiltext{1}{Current address:  School of Physics, University of New South
Wales, P.O. Box 1, Kensington, NSW 2033, AUSTRALIA}

\newpage

\begin{abstract}

  We describe our application of broad-band photometric redshift techniques to
faint galaxies in the Hubble Deep Field.  To magnitudes $AB(8140) < 26$, the
accuracy of the photometric redshifts is a few tenths and the reliability of the
photometric redshifts approaches 100\%.  At fainter magnitudes the effects of
photometric error on the photometric redshifts can be rigorously quantified and
accounted for.  We argue that broad-band photometric redshift techniques can be
applied to accurately and reliably estimate redshifts of galaxies that are up to
many magnitudes fainter than the spectroscopic limit.

\end{abstract}

\newpage

\section{INTRODUCTION}

  Between 2500 and 3500 objects (depending on exactly how they are counted) are
visible in the Hubble Deep Field (HDF) images obtained by the Hubble Space
Telescope (HST), to a limiting magnitude threshold that approaches $AB(8140)
\approx 30$.  Spectroscopic redshift identifications of just over 100 of these
objects---or roughly 3\% of the total---have been obtained with the Keck
telescope, following nearly two years of intensive effort.  The faintest objects
that have been (or presumably can be) spectroscopically identified with the Keck
telescope are of magnitude $AB(8140) \approx 26$, of which there are another 200
or so that remain unidentified in the HDF.  In time, spectroscopic redshift
identifications of perhaps 300 objects---or roughly 10\% of the total---might be
obtained with the Keck telescope.  What of the other 90\%---the objects fainter
than the spectroscopic limit?

  Fortunately, it appears to be the case that broad-band photometric redshift
techniques can be applied to accurately and reliably estimate redshifts of
galaxies that are up to many magnitudes fainter than the spectroscopic limit. 
At low redshifts, the most prominent broad-band spectral feature is the 4000
\AA\ spectral discontinuity, while at high redshifts the most prominent
broad-band spectral feature is the Lyman spectral discontinuity, which is
produced by intrinsic and intervening neutral hydrogen absorption.  But the
broad-band photometric redshift techniques make use not only of discrete
spectral discontinuities but also of overall galaxy spectral energy
distributions and so are applicable to galaxies that do not exhibit prominent
broad-band spectral features (e.g.\ low-redshift, late-type galaxies) as well as
to galaxies that do exhibit prominent broad-band spectral features (e.g.\
low-redshift, early-type galaxies or high-redshift galaxies).  To magnitudes
$AB(8140) \apl 26$, at which photometric redshifts can be compared with
spectroscopic redshifts, the accuracy of the photometric redshifts is a few
tenths and the reliability of the photometric redshifts approaches 100\%.  To
even fainter magnitudes, at which numerical simulations can quantify the effects
of photometric error, the accuracy and reliability of the photometric redshifts
are plausibly sufficient to establish statistical moments of the galaxy
distribution versus redshift, even to magnitudes close to the limiting magnitude
thresholds of the images.

  Interest in broad-band photometric redshift techniques has in the past been
driven by the prospect of determining redshifts of galaxies at relatively low
cost.  Our interest is instead driven by the prospect of determining redshifts
of galaxies that are inaccessible to ground-based spectroscopy at any cost,
i.e.\ galaxies of very low luminosity or very high redshift.  To bring this
about, it is necessary first to make the case for the photometric redshifts. 
This is, of course, a difficult agenda, because the aim is to establish the
validity of the photometric redshifts in a regime in which they by definition
cannot be independently verified.  Yet we believe this is also a crucially
important agenda.  The truly phenomenal progress toward observing normal
galaxies at high redshifts made over the past several years has simultaneously
revealed the possibilities and the limitations of ground-based spectroscopy with
the Keck telescope.  It is now clear that the very most luminous galaxies at
redshifts $z \apl 4$ are accessible to Keck spectroscopy, but it is also now
clear that lower-luminosity galaxies at redshifts much above $z \apl 4$ are
probably not accessible to Keck spectroscopy.  To push toward lower 
luminosities and higher redshifts it will almost certainly be necessary to rely
on broad-band photometric redshift techniques, perhaps supplemented by
occasional spectra of very faint objects obtained with HST (which for faint,
spatially unresolved objects at near-infrared and infrared wavelengths is a far
more sensitive spectroscopic instrument than the Keck telescope).

  The goal of this review is to make the case for the photometric redshifts.
Happily, our task is made more manageable by recent progress on two fronts: 
First, at least six independent groups (besides ours) are estimating photometric
redshifts of galaxies in the HDF---and are obtaining results that are in
remarkable agreement with each other.  (Comparisons of results of these
independent analyses are presented by Ellis 1997 and Hogg et al.\ 1997.) 
Second, at least six independent groups are obtaining spectroscopic redshifts of
galaxies in the HDF for comparison with the photometric redshifts.  Although we
concentrate on our own photometric analysis, much of what we report should apply
to the various other photometric analyses, which are similar to our analysis in
spirit if not in detail.  We begin by briefly describing the technique in \S\ 2
and the accuracy and reliability of the photometric redshifts in \S\ 3.  We
continue by applying the broad-band photometric redshift technique to galaxies
of redshift $z = 5 - 7$ in \S\ 4 galaxies of redshift $z = 7 - 17$ in \S\ 5 and
to the redshift distribution of faint galaxies in \S\ 6.  We conclude by
discussing possible future directions in \S\ 7.

\section{TECHNIQUE}

\subsection{Overview}

  Two conceptually different approaches have been applied to estimate
photometric redshifts of galaxies in the HDF.  The first technique, which we
designate the ``spectral template technique,'' is to seek optimal agreement
between the observed photometry and redshifted spectral templates.  The spectral
templates are constructed from observations of nearby galaxies or stellar
population synthesis models and from observations of intervening neutral
hydrogen absorption (measured by means of QSO absorption lines).  The second
technique, which we designate the ``empirical technique,'' is to seek optimal
agreement between the observed photometry and polynomial fits to empirical
correlations between fluxes, colors, and redshifts.  The empirical correlations
are established and calibrated with respect to known redshifts of
spectroscopically identified galaxies.  Practitioners of the two techniques are
listed in Table 1.

  The two techniques are subject to certain strengths and weaknesses as
follows:  

  1.  A strength of the spectral template technique is that it does not depend
on the spectroscopic redshifts, whereas a weakness of the empirical technique is
that it does depend on the spectroscopic redshifts.  This is an issue because
some fraction of the published spectroscopic redshifts are in error, and while
the spectral template technique provides an independent check of the
spectroscopic redshifts, the empirical technique can propagate errors in the
spectroscopic redshifts into errors in the photometric redshifts.

  2.  A strength of the spectral template technique is that it can be applied to
arbitrarily large redshifts, whereas a weakness of the empirical technique is
that it cannot be applied to arbitrarily large redshifts.

  3.  A strength of the empirical technique is that it accounts for evolution of
galaxy spectra with redshift, whereas a weakness of the spectral template
technique is that it does not account for evolution of galaxy spectra with
redshift (although the analysis of Gwyn \& Hartwick 1996 does attempt to model
time-evolving spectral templates).

  Despite the difference inherent in the techniques, there seems to be general
consensus that both techniques, properly applied, yield similar results over
the common redshift interval at which they can be compared (cf.\ Ellis 1997;
Hogg et al.\ 1997).

  In the following sections, we describe our implementation of the spectral
template technique, which is described in more detail by Lanzetta, Yahil, \&
Fern\'andez-Soto (1996), Lanzetta, Fern\'andez-Soto, \& Yahil (1997), and
Fern\'andez-Soto, Lanzetta, \& Yahil (1997, in preparation).  Specifically, we
describe our methods of (1) photometry and (2) redshift estimation, which are in
principle the only aspects of the analysis upon which results of the broad-band
photometric redshift techniques can depend.
 
\subsection{Photometry}

  Our current analysis is based on the Version 2 optical images obtained by HST
in December, 1995 using the Wide Field Planetary Camera 2 (WFPC2) and the F300W,
F450W, F606W, and F814W filters (Williams et al.\ 1996) and on the Version 1
infrared images obtained by the Kitt Peak National Observatory  (KPNO) 4 m
telescope in April, 1996 using the IRIM camera and standard $J$, $H$, and $K$
filters (Dickinson et al.\ 1997, in preparation).  First, we identify objects in
the F814W image using the SExtract source extraction program (Bertin \& Arnouts
1996).  Next, we measure the noise characteristics of the images by calculating
empirical covariances, excluding pixels associated with identified objects. 
Next, we use non-overlapping isophotal aperture masks determined by the SExtract
program to measure fluxes and flux uncertainties, applying different methods in
the optical and infrared images.

  To measure fluxes and flux uncertainties in the optical images, which are
characterized by point spread functions of ${\rm FWHM} \approx 0.1$ arcsec, we
directly integrate within the aperture mask of every object detected in the
F814W image.  To measure fluxes and flux uncertainties in the infrared images,
which are characterized by point spread functions of ${\rm FWHM} \approx 1$
arcsec, we (1) model the spatial profile of every object detected in the F814W
image as a convolution of the portion of the F814W image containing the object
with the appropriate point spread function of the infrared image and (2)
determine a least-squares fit of a linear sum of the model spatial profiles to
the infrared image.  The advantages of this method over simple aperture
photometry are that (1) the flux measurements correctly weight signal-to-noise
ratio variations within the spatial profiles and (2) the flux uncertainty
measurements correctly include the contributions of nearby, overlapping
neighbors.  The observed and modeled $K$-band images are shown in Figure 1.

  Our analysis of the infrared images yields essentially optimal photometry
under the assumption that the spatial profiles of the objects are independent of
wavelength at observed-frame wavelengths $\lambda > 8140$ \AA.  This assumption
may be justified as follows:  At low redshifts, the infrared images measure
rest-frame optical and infrared wavelengths, at which galaxy spatial profiles
are observed to be more or less independent of wavelength.   At high redshifts,
galaxies are (as it turns out) physically small and unresolved by the infrared
images at any wavelength.  A similar analysis of deep ground-based optical
images of the HDF (e.g.\ through medium-band filters) could certainly be applied
to great advantage, making use of the unprecedented depth and spatial resolution
of the F814W image to provide nearly ideal photometric spatial templates.

\subsection{Redshift Estimation}

  First, we construct spectral templates of E/S0, Sbc, Scd, and Irr galaxies,
including the effects of intrinsic and intervening neutral hydrogen absorption.
Next, we integrate the redshifted spectral templates with the throughputs of
the F300W, F450W, F606W, F814W, $J$, $H$, and $K$ filters, at redshifts spanning
$z = 0 - 7$.   Next, we construct the ``redshift likelihood function'' of every
object detected in the F814W image by calculating the relative likelihood of
obtaining the measured fluxes and uncertainties given the modeled fluxes at an
assumed redshift, maximizing with respect to galaxy spectral type and arbitrary
flux normalization.  Finally, we determine the maximum-likelihood redshift
estimate of every object detected in the F814W image by maximizing the redshift
likelihood function with respect to redshift.

  The spectral templates are constructed from different sources at
far-ultraviolet, near-ultraviolet and optical, and infrared wavelengths. 
First, we adopt empirical spectra of Coleman, Wu, \& Weedman (1980) at
wavelengths $\lambda = 1400 - 10,000$ \AA.  Next, we add spectra derived from
simple power-law relationships motivated by results of Kinney et al.\ (1993) at
wavelengths $\lambda = 912 - 1400$ \AA.  Next, we add spectra derived from
stellar population synthesis models of Bruzual \& Charlot (1993) at wavelengths
$\lambda = 10,000 - 25,000$ \AA.  Finally, we include the effects intrinsic and
intervening neutral hydrogen absorption (as a function of redshift) by (1)
assuming that galaxies are optically thick at wavelengths shortward of the Lyman
limit and (2) adding absorption from intervening \lya-forest \lya\ and \lyb\
absorption lines according to the ``flux decrement'' parameters $D_A$ and 
$D_B$, measured by means of QSO absorption lines by Madau (1995) and Webb
(1996, unpublished).  The spectral templates are shown in Figure 2.

  We experimented with spectral templates based at near-ultraviolet and optical
wavelengths on (1) empirical spectra of Kinney et al.\ (1996) and (2) spectra
derived from stellar population synthesis models of Bruzual \& Charlot (1993). 
We abandoned these alternate spectral templates in favor of the adopted 
spectral templates because photometric redshifts determined with the alternate
spectral templates did not compare as favorably with the spectroscopic redshifts
as photometric redshifts determined with the adopted spectral templates.

\section{ACCURACY AND RELIABILITY OF PHOTOMETRIC REDSHIFTS}

  In this section we assess the accuracy and reliability of the photometric
redshifts by (1) direct comparison of photometric and spectroscopic redshifts
and (2) numerical simulations of the effects of photometric error on the
photometric redshifts.

\subsection{Photometric Redshifts:  Successes and Failures}

  In preparation for the Hubble Deep Field Symposium, we set out in early May,
1997 to compare the available photometric redshifts with the available
spectroscopic redshifts, which at the time numbered roughly 80 redshifts
obtained with the Keck telescope.  To our dismay, we found that a surprisingly
large fraction (roughly 15\%) of the photometric redshifts were clearly
discordant with the spectroscopic redshifts, by amounts that were far too large
to be attributed to photometric error.  Believing that at the relatively bright
magnitudes of the spectroscopic limit the reliability of the photometric
redshifts should in principle be much higher than suggested by this comparison,
we examined each of the discordant redshift pairs in detail, expecting to learn
what caused the photometric redshifts to fail.  Instead we found that in nearly
every case it was the spectroscopic redshift---not the photometric
redshift---that was in error.

  That some fraction of the published spectroscopic redshifts are in error is
hardly unexpected.  After all, determining redshifts of extremely faint
galaxies by any means---photometric or spec\-troscopic---is a difficult process
subject to a variety of systematic uncertainties.  But an uncritical acceptance
of the validity of the spectroscopic redshifts has led some previous authors to 
incorrectly (and negatively) assess the reliability of the photometric
redshifts.  For this reason, it is important that any comparison of the
photometric and spectroscopic redshifts begin by establishing under what
circumstances both the photometric redshifts and the spectroscopic redshifts
fail to yield reliable results.  Our comparison of the photometric and
spectroscopic redshifts in early May, 1997 identified eight spectroscopic
redshifts that are ambiguous or in error and three photometric redshifts that
are in error.  (In two cases, both the spectroscopic and the photometric
redshift are in error.)  Here we describe each of the discordant cases,
identifying galaxies by the Version 2 $x$ and $y$ mosaic pixel coordinates.

\subsubsection{The Successes}

  {\em Galaxy 1591,3681:}  This galaxy was assigned a spectroscopic redshift of
$z = 0.13$ by Cowie (1997) and a photometric redshift of $z = 1.20$ by our
analysis.  The galaxy occurs within $\approx 1$ arcsec of a slightly brighter
galaxy with no available spectroscopic redshift and a photometric redshift of $z
= 0.16$.  Considering typical ground-based seeing, it is likely that the
spectroscopic redshift refers to the nearby galaxy, of which the photometric
redshift is in excellent agreement with the spectroscopic redshift.  In this
case, the spectroscopic redshift is ambiguous or in error due to confusion with
an overlapping source.

  {\em Object 2449,1574:}  This galaxy was assigned a spectroscopic redshift
of $z = 0.318$ by Cowie (1997) and a photometric redshift of $z = 4.48$ by our
analysis.  The object occurs within $\approx 2$ arcsec of a much brighter galaxy
with a spectroscopic redshift of $z = 0.321$ and a photometric redshift of $z =
0.24$.  The published spectra of the two objects are virtually identical
although the colors of the two galaxies are very dissimilar, suggesting that one
or the other or both of the spectra are incorrectly assigned.  Subsequent
spectroscopic observations of this object have established that it is a star
(Cowie 1997, private communication).  In this case, the spectroscopic redshift
is in error due to operator error.  This object is also noted below as a
failure, because the photometric redshift is discordant with the actual
redshift.

  {\em Galaxy 561,305:}  This galaxy was assigned a spectroscopic redshift
of $z = 0.47$ by Cowie (1997) and a photometric redshift of $z = 0.76$ by our
analysis.  The galaxy occurs within $\approx 1$ arcsec of a slightly brighter
galaxy with no available spectroscopic redshift and a photometric redshift of
$z = 0.48$.  As in the case of galaxy 1591,3681, it is likely that the
spectroscopic redshift refers to the nearby galaxy, of which the photometric
redshift is in excellent agreement with the spectroscopic redshift.  In this
case, the spectroscopic redshift is ambiguous or in error due to confusion with
an overlapping source.

  {\em Galaxies 790,1877 and 745,1845:}  These galaxies were assigned
spectroscopic redshifts of $z = 0.511$ (Cowie 1997) and photometric redshifts of
$z = 2.20$ (galaxy 790,1877) and $z = 0.56$ (galaxy 745,1845).  The galaxies
occur within $\approx 3$ arcsec of each other.  The published spectra of the two
galaxies are identical, indicating that one or the other or both of the spectra
are incorrectly assigned.  In this case, the spectroscopic redshifts are
ambiguous or in error due to operator error.

  {\em Galaxy 693,1224:}  This galaxy was assigned a spectroscopic redshift of
$z = 1.231$ by Cowie (1997) and a photometric redshift of $z = 1.08$ by our
analysis.  The galaxy occurs within $\approx 1$ arcsec of a galaxy with no
available spectroscopic redshift and a photometric redshift of $z = 1.18$.  As
in the case of galaxy 1591,3681, it is likely that the spectroscopic redshift
refers to the nearby galaxy, of which the photometric redshift is in excellent
agreement with the spectroscopic redshift.  In this case, the spectroscopic
redshift is ambiguous or in error due to confusion with an overlapping source.

  {\em Galaxy 1816,1026:}  This galaxy was assigned a spectroscopic redshift of
$z = 2.775$ by Steidel et al.\ (1996) and a photometric redshift of $z = 1.72$
by our analysis.  The observed photometry of this galaxy is compared with the
model spectrum of an Irr galaxy of redshift $z = 2.775$ in Figure 3.  Results of
Figure 3 indicate that the spectroscopic redshift cannot be correct unless our
model of intrinsic and intervening neutral hydrogen is grossly incorrect. 
Furthermore, our examination of the published spectrum of this galaxy (which
is of moderate signal-to-noise ratio) fails to discern the claimed absorption
features upon which the spectroscopic redshift is based.  Subsequent analysis
of the spectrum of this galaxy has failed to verity the spectroscopic redshift
(Steidel 1997, private communication).  In this case, the spectroscopic redshift
is ambiguous or in error due to inaccurate identification of spectral features.

  {\em Galaxy 1193,3751:}  This galaxy was assigned a spectroscopic redshift of
$z = 2.845$ by Steidel et al.\ (1996) and a photometric redshift of $z = 0.04$
by our analysis.  The observed photometry of this galaxy is compared with the
model spectrum of an Irr galaxy of redshift $z = 2.845$ in Figure 4.  As in the
case of galaxy 1816,1026, results of Figure 4 indicate that the spectroscopic
redshift cannot be correct unless our model of intrinsic and intervening neutral
hydrogen is grossly incorrect, and our examination of the published spectrum of
this galaxy (which is of only moderate signal-to-noise ratio) fails to discern
the claimed absorption features upon which the spectroscopic redshift is based.
Subsequent spectroscopic observations of this galaxy have established a
spectroscopic redshift of $z = 2.008$ (Steidel 1997, private communication).  In
this case, the spectroscopic redshift is in error due to inaccurate
identification of spectral features.  This galaxy is also noted below as a
failure, because the photometric redshift is discordant with the actual
redshift.

\subsubsection{The Failures}

  {\em Object 2449,1574:}  This object was assigned a photometric redshift of
$z = 4.48$ by our analysis and a spectroscopic redshift of $z = 0.318$ by Cowie
(1997), although as noted above subsequent spectroscopic observations of this
object have established that it is a star (Cowie 1997, private communication).
Our analysis does not include spectral templates of stars, which therefore must
be removed by hand.  In this case, the photometric redshift is in error due to
confusion by a star.

  {\em Galaxy 1193,3751:}  This galaxy was assigned a photometric redshift of
$z = 0.04$ by our analysis and a spectroscopic redshift of $z = 2.845$ by
Steidel et al.\ (1996), although as noted above subsequent spectroscopic
observations of this galaxy have established a spectroscopic redshift of $z =
2.008$ (Steidel 1997, private communication).  The redshift likelihood function
of this galaxy shows no local maxima near the actual redshift $z \approx
2.008$.  In this case, the photometric redshift is in error due to cosmic
variance with respect to the spectral templates.  This galaxy is also noted
above as a success, because the photometric analysis demonstrates that the
original spectroscopic redshift cannot be correct.

  {\em Galaxy 3242,1972:}  This galaxy was assigned a photometric redshift of 
$z = 0.00$ by our analysis and a spectroscopic redshift of $z = 2.591$ by
Steidel et al.\ (1996).  The redshift likelihood function of this galaxy shows
a local maximum at redshift $z = 2.00$, which has a likelihood $2.9 \sigma$
less than the global maximum at $z = 0.00$.  In this case, the photometric
redshift is in error due to cosmic variance with respect to the spectral
templates.

  {\em Galaxy 2888,1503:}  This galaxy was assigned a photometric redshift of
$z = 1.32$ by our analysis and a spectroscopic redshift of $z = 2.268$ by
Steidel et al.\ (1996).  The redshift likelihood function of this galaxy shows
no local maxima near the actual redshift $z \approx 2.268$.  In this case, the
photometric redshift is in error due to cosmic variance with respect to the
spectral templates.

\subsubsection{Summary}

  Spectroscopic redshifts of faint galaxies in the HDF are subject to a
misidentification rate of $\approx 10\%$, which results from operator error,
confusion with overlapping sources, and inaccurate identification of spectral
features.  Photometric redshifts of faint galaxies in the HDF are subject to a
misidentification rate of $\approx 5\%$, which results from cosmic variance with
respect to the spectral templates and confusion with stars.

\subsection{Comparison of Photometric and Spectroscopic Redshifts}

  Spectroscopic redshift identifications of nearly 120 objects have been
obtained with the Keck telescope as of August, 1997.  Magnitudes of the
spectroscopically identified objects range from $AB(8140) \approx 18$ through
$AB(8140) \approx 26$, and redshifts of the spectroscopically identified 
objects range from $z \approx 0$ through $z \approx 4$.  Excluding from
consideration (1) spectroscopic redshifts that are described as ambiguous or in
error in \S\ 3.1.1 or as uncertain by the spectroscopic observers and (2)
spectroscopic redshifts of stars (for which our analysis does not include
spectral templates), spectroscopic redshifts of 102 galaxies are available for
comparison with the photometric redshifts.  Sources of the spectroscopic
redshifts are listed in Table 3, and the comparison of photometric and
spectroscopic redshifts is shown in Figure 5.  Results of Figure 5 indicate that
the photometric redshifts are in broad agreement with the spectroscopic
redshifts.  At redshifts $z < 2$, the photometric redshifts generally trace the
spectroscopic redshifts, with no examples of photometric redshifts that are
clearly discordant with the spectroscopic redshifts.  At redshifts $z > 2$, the
photometric redshifts generally trace the spectroscopic redshifts, with two
examples of photometric redshifts that are clearly discordant with the
spectroscopic redshifts.  (The discordant redshifts arise from galaxies
1193,3751 and 3242,1972, which are discussed in \S\ 3.1.2.  In both cases, a
high-redshift galaxy is incorrectly assigned a low redshift.)

  To quantitatively assess the accuracy and reliability of the photometric
redshifts by direct comparison of the photometric and spectroscopic redshifts,
it is necessary to adopt some robust measure of the distribution of residuals
between the photometric and spectroscopic redshifts.  This is because (1) the
residuals are not drawn from a Gaussian distribution and (2) the RMS residual is
dominated by a few highly discordant redshifts, as is evident from Figure 5. 
Here we consider two such measures of the residuals:  First, we consider the
median absolute residual between the photometric and spectroscopic redshifts. 
Next, we  consider the ``clipped'' RMS residual between the photometric and
spectroscopic redshifts.  The clipped RMS residual is obtained by applying an
iterative ``sigma-clipping'' algorithm to the residuals, rejecting as discordant
photometric redshifts for which the residual exceeds three times the clipped RMS
residual.  According to this procedure, the ``accuracy'' of the photometric
redshifts is measured by the clipped RMS residual and the ``reliability'' of the
photometric redshifts is measured by the discordant fraction.  Results are
summarized in Table 3, which lists the sample condition, the median absolute
residual, the clipped RMS residual, the number of discordant redshifts, the
number of total redshifts, and the discordant fraction.

  Several conclusions can be drawn from results of Table 3:

  1.  At the redshifts $z \approx 0 - 4$ spanned by the comparison of
photometric and spectroscopic redshifts, the accuracy of the photometric
redshifts is a few tenths and the reliability of the photometric redshifts
approaches 100\%.  Specifically, the median absolute residual is 0.09, the
clipped RMS residual is 0.17, and the discordant fraction is 0.049. 

  2.  The accuracy of the photometric redshifts is greater at low redshifts than
at high redshifts, but the {\em fractional} accuracy of the photometric
redshifts is greater at high redshifts than at low redshifts.  At redshifts $z <
2$, the clipped RMS residual is 0.09, which at a representative redshift $z =
0.5$ corresponds to a fractional accuracy of $\approx 18\%$.  At redshifts $z >
2$, the clipped RMS residual is 0.40, which at a representative redshift $z = 3$
corresponds to a fractional accuracy of $\approx 13\%$.

  3.  The accuracy of the photometric redshifts is greater for late-type
galaxies (which do not exhibit prominent broad-band spectral features) than for
early-type galaxies (which do exhibit prominent broad-band spectral features). 
At redshifts $z < 2$ (at which early-type galaxies are found), the clipped RMS
residual for late-type galaxies is 0.09 and the clipped RMS residual for
early-type galaxies is 0.25.

  There is some evidence that the photometric redshifts of galaxies at
redshifts $z \apg 2$ are systematically underestimated by $\approx 0.3$,
especially at redshifts $z = 2 - 3$.  If this systematic offset is accounted
for, at redshifts $z > 2$ the clipped RMS residual is decreased to 0.24 and the
discordant fraction is increased to 0.115.  We have tried but failed to
determine the cause of this systematic offset.

\subsection{Photometric Error versus Cosmic Variance}

  There are in principle two effects that contribute to residuals between the
photometric and spectroscopic redshifts: photometric error and cosmic variance
with respect to the spectral templates.  Here we evaluate the relative
importance of photometric error and cosmic variance by means of a numerical
simulation of the effect of photometric error on the photometric redshifts. 
First, we add random noise (according to the actual flux uncertainties) to the
best-fit model fluxes of every object detected in the F814W image.  Next, we
redetermine the maximum-likelihood redshift estimate of every object detected in
the F814W image, using the simulated rather than measured fluxes.  Finally, we
repeat these steps 100 times in order to determine distributions of residuals
between the actual and simulated redshifts.  Results are summarized in Figure 6,
which shows the distributions of residuals and the median absolute residuals
between the actual and simulated redshifts as functions of magnitude and
redshift.

  Several conclusions can be drawn from results of Figure 6:

  1.  Photometric error produces almost no effect on the photometric redshifts
at magnitudes $AB(8140) < 25$.  At these magnitudes, the median absolute
residual is no greater than 0.04, and the distributions of residuals are
concentrated into single peaks centered at zero residual.

  2.  Photometric error produces a mild effect on the photometric redshifts at
magnitudes $AB(8140) = 25 - 26$, which is somewhat more pronounced at redshifts
$z > 2$ than at redshifts $z < 2$.  Although at these magnitudes the median
absolute residual (at all redshifts) is only 0.04 (which is comparable to the
median absolute residual at brighter magnitudes), the distributions of residuals
at high-redshifts are spread into small secondary peaks centered at large
negative residuals in addition to prominent primary peaks centered at zero
residual.  These secondary peaks---which result from high-redshift galaxies that
are incorrectly assigned low redshifts---contain 2.5\% of the total at redshifts
$z = 2 - 3$ and 6.1\% of the total at redshifts $z = 3 - 4$.

  3.  Photometric error produces an increasingly significant effect on the
photometric redshifts with increasing magnitude at magnitudes $AB(8140) > 26$. 
At redshifts $z = 0 - 1$, the effect of photometric error at faint magnitudes is
to (1) increase the median absolute residuals and (2) produce long tails
stretching to large positive residuals on the distributions of residuals.  At
redshifts $z > 1$, the effect of photometric error at faint magnitudes is to (1)
increase the median absolute residuals and (2) produce prominent secondary peaks
centered at large negative residuals on the distributions of residuals.  At
magnitudes $AB(8140) = 26 - 27$, these secondary peaks---which again result from
high-redshift galaxies that are incorrectly assigned low redshifts---contain
9.7\% of the total at redshifts $z = 2 - 3$ and 19.6\% of the total at redshifts
$z = 3 - 4$.

  4.  At faint magnitudes, it is more likely that high-redshift galaxies will be
incorrectly assigned low redshifts than that low-redshift galaxies will be
incorrectly assigned high redshifts.  For example, at magnitudes $AB(8140) =
27 - 28$, the long tail stretching to large positive residuals at redshifts $z =
0 - 1$ contains 27.3\% of the total while the secondary peak at large negative
residuals at redshifts $z = 3 - 4$ contains 34.8\% of the total.

  It is possible to drawn general conclusions about the relative importance of
photometric error and cosmic variance on the photometric redshifts by combining
results of Figure 6 with results of \S\ 3.2.  At redshifts $z < 2$, magnitudes
of the spectroscopically identified objects range from $AB(8140) = 18.16$
through $AB(8140) =  24.88$, with a median of $AB(8140) = 22.68$.  Results of
Figure 6 indicate that at these magnitudes photometric error produces almost no
effect on the photometric redshifts.  At redshifts $z > 2$, magnitudes of the
spectroscopically identified objects range from $AB(8140) = 23.40$ through
$AB(8140) =  26.75$ with a median of $AB(8140) = 25.08$.  Results of Figure 6
indicate that at these magnitudes photometric error produces at most a mild
effect on the photometric redshifts.  [For example, even at magnitudes as faint
as $AB(8140) = 26 - 27$, the median absolute residual produced by cosmic
variance is less than the median absolute residual between the photometric and
spectroscopic redshifts, although the discordant fraction produced by cosmic
variance is consistent with the discordant fraction between the photometric and
spectroscopic redshifts.]  We therefore conclude that {\em residuals between the
photometric and spectroscopic redshifts are dominated by cosmic variance rather
than by photometric error}.

  This result establishes the magnitude of the effect of cosmic variance on the
photometric redshifts.  Specifically, by attributing the entire residual between
the photometric and spectroscopic redshifts to the effect of cosmic variance and
considering the redshifts $z \approx 0 - 4$ spanned by the comparison of
photometric and spectroscopic redshifts, we conclude that cosmic variance
produces a median absolute residual of 0.09, a clipped RMS residual of 0.17, 
and a discordant fraction of $0.049$.  As long as the effect of cosmic variance
does not increase with decreasing galaxy luminosity---and we can think of no
reason that it should---then {\em these values must apply at all magnitudes},
not just the relatively bright magnitudes of the spectroscopic limit.  In
practice, this has two important implications:  First, at magnitudes brighter
than $AB(8140) \approx 26 - 27$ the accuracy and reliability of the photometric
redshifts is limited by cosmic variance and at magnitudes fainter than $AB(8140)
\approx 26 -27$ the accuracy and reliability of the photometric redshifts is
limited by photometric error.  Second, the uncertainty associated with any
statistical moment of the galaxy distribution---including the effects of
sampling error, photometric error, and cosmic variance---may be realistically
estimated by means of a ``bootstrap'' resampling technique, simulating the
effect of photometric error by adding random noise to the fluxes and simulating
the effect of cosmic variance by adding (rather modest) random noise to the
estimated redshifts.  We expect that this technique will ultimately play an
important role in exploiting the full potential of the broad-band photometric
redshift techniques.

\section{GALAXIES OF REDSHIFT $z = 5 - 7$}

  In this section we describe an application of the broad-band photometric
redshift techniques to galaxies of redshift $z = 5 - 7$.  A full description of
this analysis is reported by Lanzetta, Fern\'andez-Soto, \& Yahil (1997).

  Results of \S\ 3 demonstrate the accuracy and reliability of the broad-band
photometric redshift techniques at the redshifts $z \approx 0 - 4$ spanned by
the comparison of photometric and spectroscopic redshifts.  But our analysis
identifies galaxies with estimated redshifts as large as $z \approx 5 - 7$,
which are probably inaccessible to ground-based spectroscopy.  These galaxies
are characterized by strong flux at observed-frame wavelengths $\lambda \approx
8140$ \AA\ and an absence of detectable flux at shorter wavelengths, hence it is
not possible by means of optical observations alone to establish continuum
spectral energy distributions of the galaxies or to definitively rule out the
competing alternatives that (1) the flux detected at observed-frame wavelengths
$\lambda \approx 8140$ \AA\ arises due to strong emission lines or (2) the
absence of detectable flux at shorter wavelengths arises due to reddening by
dust.

  To address these issues, we consider the infrared images obtained with the
KPNO 4 m telescope by Dickinson et al.\ (1997, in preparation).  Specifically,
we concentrate on galaxy 2342,968, which is the brightest galaxy with estimated
redshift $z > 5$ (at least at 8140 \AA).  The infrared photometric analysis
applied to galaxy 2342,968 is shown in Figure 7, and the spectral energy
distribution of galaxy 2342,968 is shown in Figure 8.  Results of Figures 7 and
8 indicate that galaxy 2342,968 is detected at the $0.9 \sigma$ level of
significance at the $J$ band, the $2.7 \sigma$ level of significance at the $H$
band, and the $4.7 \sigma$ level of significance at the $K$ band.  The primary
conclusions of the analysis are two-fold:  First, the detection of galaxy
2342,968 at the $J$, $H$, and $K$ infrared bands establishes the continuum
spectral energy distribution at observed-frame wavelengths spanning $\lambda
\approx 8140$ \AA\ through $\lambda \approx 2.2$ $\mu$.  This rules out the
possibility that galaxy 2342,968 is a lower-redshift galaxy that exhibits a
strong emission line at observed-frame wavelength $\lambda \approx 8140$ \AA. 
Second, the detection of galaxy 2342,968 at modest flux levels at the $J$, $H$,
and $K$ infrared bands limits the permitted reddening.  This rules out the
possibility that galaxy 2342,968 is a lower-redshift galaxy that is heavily
obscured and reddened by dust.

  To establish a quantitative measure of the degree to which heavily obscured
and reddened lower-redshift galaxies can be ruled out, we repeat the
maximum-likelihood redshift determinations of our previous analysis, modifying
the galaxy spectral templates by a grid of dust extinction models with color
excess $E(B-V)$ ranging from $E(B-V) = 0$ (no dust obscuration) through $E(B-V)
= 4$ (very heavy dust obscuration).  It is possible to find acceptable 
high-redshift solutions with very modest amounts of dust extinction, but it is
not possible to find acceptable low-redshift solutions with larger amounts of
dust extinction.  In particular, any solution with $z < 4.4$ is ruled out at
more than the $4 \sigma$ level of significance.  The reason is that it is not
possible to simultaneously satisfy the sharp discontinuity in flux between the
F606W and F814W filters and the absence of strong infrared flux with any
low-redshift galaxy spectral template and any amount of dust extinction.  A
similar analysis applied to the other galaxies with estimated redshift $z > 5$
yields similar results, although with less stringent limits.  We conclude that
the available evidence supports the photometric redshift identifications and
that it is very unlikely that at least the brightest of the galaxies with
estimated redshifts $z > 5$ are at substantially lower redshifts. 

\section{REDSHIFT DISTRIBUTION OF FAINT GALAXIES}

  Our analysis determines photometric redshifts of over 2500 galaxies, which
together with an assessment of the effects of photometric error and cosmic
variance can be used to construct an essentially exhaustive account of
statistical properties (and uncertainties) of faint galaxies at magnitudes
ranging to $AB(8140) \approx 30$ and redshifts ranging to $z \approx7$.  A
complete enumeration of the properties of faint galaxies is beyond the scope of
this review, so we instead illustrate the utility of broad-band photometric
redshift techniques by presenting the redshift distribution of faint galaxies. 
The redshift distribution and median redshift of faint galaxies as functions of
limiting magnitude threshold are shown in Figure 9:

  Several conclusions can be drawn from results of Figure 9:

  1.  At the brightest magnitudes $AB(8140) \approx 23$ accessible to the HDF,
the galaxy redshift distribution is peaked at low redshifts with a median
redshift of $z = 0.64$.  This agrees with results of deep ground-based galaxy
surveys, the deepest of which are sensitive to comparable magnitudes (e.g.\
Cowie et al.\ 1996).

  2.  High-redshift galaxies with $z > 2$ become prominent at magnitudes
$AB(8140) \approx 25$, at which they constitute $\approx 6.1\%$ of the galaxy
population.  At fainter magnitudes, high-redshift galaxies become increasingly
prominent with increasing magnitude threshold.  Galaxies of redshift $z > 2$
constitute $\approx 13.9\%$ of the galaxy population at magnitudes $AB(8140) <
26$, $19.5\%$ of the galaxy population at magnitudes $AB(8140) < 27$, and
$24.0\%$ of the galaxy population at magnitudes $AB(8140) < 28$.

  4.  The median redshift of the galaxy redshift distribution increases
monotonically with increasing magnitude threshold.  The median redshift of the
galaxy redshift distribution grows from ${\rm med} \ z = 0.68$ at magnitudes
$AB(8140) < 24$, to ${\rm med} \ z = 1.04$ at magnitudes $AB(8140) < 26$, to
${\rm med} \ z = 1.36$ at magnitudes $AB(8140) < 28$.

  5.  The galaxy redshift distribution exhibits a precipitous decline at
redshift $z \approx 2$ at all magnitudes $AB(8140) \apg 26$.  This is unlikely
to be an artifact of the analysis because (1) results of \S\ 3 indicate that
redshifts $z > 2$ exhibit a comparable accuracy to redshifts $z < 2$ and (2)
there are prominent broad-band spectral features redshifted into the optical
filters at redshifts $z \approx 2$.

\section{FUTURE DIRECTIONS}

  Our conclusion is that broad-band photometric redshift techniques can be
applied to accurately and reliably estimate redshifts of galaxies that are
inaccessible to ground-based spectroscopy.  At bright magnitudes the 
reliability of the photometric redshifts compares favorably with the reliability
of the spectroscopic redshifts, and at faint magnitudes the effects of
photometric error on the photometric redshifts can be rigorously quantified and
accounted for.  Yet we suspect that the ability of broad-band photometric
techniques to reliably identify galaxies at redshifts beyond the ground-based
spectroscopic limit will not become generally accepted until some means of
independent verification can be found.  The very brightest galaxy at estimated
redshifts $z > 5$ (galaxy 2342,968) is of magnitude $AB(8140) = 26.3$, which is
probably inaccessible to Keck spectroscopy.  How, then, is progress to be
made?

  It is easy to show that for background-limited observations of spatially
unresolved objects, the ratio of exposure times needed to obtain a given
signal-to-noise ratio for space-based versus ground-based observations 
(assuming identical system throughputs and bandpasses) scales as
\begin{equation}
\frac{t_{\rm ground}}{t_{\rm space}} =
\frac{f_\nu({\rm sky})_{\rm ground}}{f_\nu({\rm sky})_{\rm space}}
\frac{A_{\rm space}}{A_{\rm ground}}
\frac{\Omega({\rm sky})_{\rm ground}}{\Omega({\rm sky})_{\rm space}},
\end{equation}
where $f_\nu({\rm sky})$ is the sky brightness, $A$ is the telescope aperture,
and $\Omega({\rm sky})$ is the solid angle subtended by a spatial resolution
element.  At near-infrared wavelengths the first term of this expression
evaluates to $\approx 4.4$, for HST versus the Keck telescope the second term
evaluates to $\approx 0.06$, and for space-based seeing of $\approx 0.1$ arcsec
versus (optimistic) ground-based seeing of $\approx 0.6$ the third term
evaluates to $\approx 36$.  Forming the product of these three terms, we
conclude that {\em HST is roughly 10 times more sensitive than the Keck
telescope for spectroscopy of faint, spatially unresolved objects}.  Although
the Keck telescope can take advantage of multi-object spectroscopic
capabilities---which HST with STIS probably cannot---this is not an issue for
objects at the very limit of detectability or for objects of very low spatial
density.  We believe that to establish the broad-band photometric techniques at
magnitudes and redshifts beyond the spectroscopic limit it will be necessary
to use HST with STIS obtain spectroscopic confirmation of a few targets
chosen carefully on the basis of broad-band photometric redshifts.  These
observations will provide the means of establishing the effects of cosmic
variance on the photometric redshifts at redshifts $z \apg 4$.

\acknowledgements

  We are grateful to R.\ Williams and the staff of STScI for providing access
to the reduced optical images, to M.\ Dickinson for providing access to the
reduced infrared images, and to C.\ Steidel and M.\ Dickinson providing access
to spectroscopic redshifts in advance of publication.  A.F.-S. and K.M.L. were
supported by NASA grant NAGW--4422 and NSF grant AST--9624216.

\newpage

\begin{center}
\begin{tabular}{p{1.5in}cc}
\multicolumn{3}{c}{TABLE 1} \\
\multicolumn{3}{c}{PRACTITIONERS OF BROAD-BAND} \\
\multicolumn{3}{c}{PHOTOMETRIC REDSHIFT TECHNIQUES} \\
\hline
\multicolumn{1}{c}{Group} & Technique & Reference \\
\hline
\hline
Hawaii \dotfill           & spectral template & 1 \\
Imperial College \dotfill & spectral template & 2 \\
Johns Hopkins \dotfill    & empirical         & 3 \\
Stony Brook \dotfill      & spectral template & 4 \\
Toronto \dotfill          & spectral template & 5 \\
Victoria \dotfill         & spectral template & 6 \\
\hline
\multicolumn{3}{p{3.65in}}{REFERENCES---(1) Cowie 1997; (2) Mobasher et al. 1996;
(3) Connolly et al.\ 1997; (4) Lanzetta, Yahil, \& Fern\'andez-Soto 1996; (5)
Sawicki et al.\ 1997; (6) Gwyn \& Hartwick 1996.}
\end{tabular}
\end{center}

\newpage

\begin{center}
\begin{tabular}{p{2in}c}
\multicolumn{2}{c}{TABLE 2} \\
\multicolumn{2}{c}{SOURCES OF SPECTROSCOPIC} \\
\multicolumn{2}{c}{REDSHIFTS} \\
\hline
\multicolumn{1}{c}{Group} & Reference \\
\hline
\hline
Berkeley \dotfill         & 1 \\
Caltech \dotfill          & 2 \\
Elston \dotfill           & 3 \\
Hawaii \dotfill           & 4 \\
Lick \dotfill             & 5 \\
Steidel et al.\ \dotfill  & 6,7 \\
\hline
\multicolumn{2}{p{3.0in}}{REFERENCES---(1) Zepf, Moustakas, \& Davis 1997; (2)
Cohen et al.\ 1997; (3) Elston 1997, private communication; (4) Cowie 1997; (5)
Lowenthal et al.\ 1997; (6) Steidel et al.\ 1996; (7) Dickinson et al.\ 1997,
this volume.}
\end{tabular}
\end{center}

\newpage

\begin{center}
\begin{tabular}{p{1.75in}cccccc}
\multicolumn{7}{c}{TABLE 3} \\
\multicolumn{7}{c}{COMPARISON OF PHOTOMETRIC AND SPECTROSCOPIC REDSHIFTS} \\
\hline
& Median & & \multicolumn{4}{c}{Clipped RMS} \\
\cline{4-7}
& Absolute & & & Discordant & Total & Discordant \\
\cline{2-2}
\multicolumn{1}{c}{Condition} & Residual & & Residual & Number & Number &
Fraction \\
\hline
\hline
All \dotfill                     & 0.09 & & 0.17 & 5 & 102 & 0.049 \\
$z < 2$ \dotfill                 & 0.08 & & 0.09 & 4 &  78 & 0.051 \\
$z > 2$ \dotfill                 & 0.32 & & 0.40 & 2 &  24 & 0.083 \\
$z < 2$, ``early type'' \dotfill & 0.12 & & 0.25 & 0 &  10 & 0.000 \\
$z < 2$, ``late type'' \dotfill  & 0.07 & & 0.09 & 2 &  68 & 0.029 \\
\hline
\end{tabular}
\end{center}

\newpage

\figcaption{Observed (left panel) and modeled (right panel) $K$-band images. 
Angular extent is $x.xx \times x.xx$ arcmin$^2$, and spatial resolution is
${\rm FWHM} \approx 1$ arcsec.}

\figcaption{Spectral templates of E/S0 (solid curve), Sbc (dotted curve), Scd
(short dashed curve), and Irr (long dashed curve) galaxies.}

\figcaption{Observed photometry of galaxy 1816,1026 compared with model
spectrum of an Irr galaxy of redshift $z = 2.775$.  Vertical error bars 
indicate $1 \sigma$ uncertainties and horizontal error bars indicate filter
FWHM.  It is very unlikely that the redshift of the galaxy is $z = 2.775$.}

\figcaption{Observed photometry of galaxy 1193,3751 compared with model
spectrum of an Irr galaxy of redshift $z = 2.845$.  Vertical error bars 
indicate $1 \sigma$ uncertainties and horizontal error bars indicate filter
FWHM.  It is very unlikely that the redshift of the galaxy is $z = 2.845$.}

\figcaption{Comparison between photometric redshifts $z_{\rm phot}$ and
spectroscopic redshifts $z_{\rm spec}$.  Solid line shows $z_{\rm phot} = z_{\rm
spec}$.}

\figcaption{Distributions of residuals between best-fit model and simulated
redshifts, illustrating the effects of photometric error on the photometric
redshifts.  Panels are arranged horizontally according to magnitude, from
$AB(8140) = 21 - 22$ through $AB(8140) = 29 - 30$ in steps of 1 mag, and
vertically according to redshift, from $z = 0 - 1$ through $z = 4 - 5$ in steps
of 1.  The median absolute residual is listed at the upper right corner of every
panel.}

\figcaption{Infrared photometric analysis applied to galaxy 2342,968.  Angular
extent of each image is $20.5 \times 20.5$ arcsec$^2$.  Left panels show
$J$-band images, center panels show $H$-band images, and right panels show
$K$-band images.  Top panels show raw images, and bottom panels show images for
which models of all objects other than galaxy B have been subtracted.  Arrows
point to the expected position of galaxy B, as measured from the F814W image.}

\figcaption{Spectral energy distribution of galaxy 2342,968.  Vertical error
bars indicate flux uncertainties, and horizontal error bars indicate filter
FWHM.}

\figcaption{Redshift distribution of faint galaxies as a function of limiting
magnitude threshold.  The limiting magnitude threshold and median redshift is
listed in every panel.} 

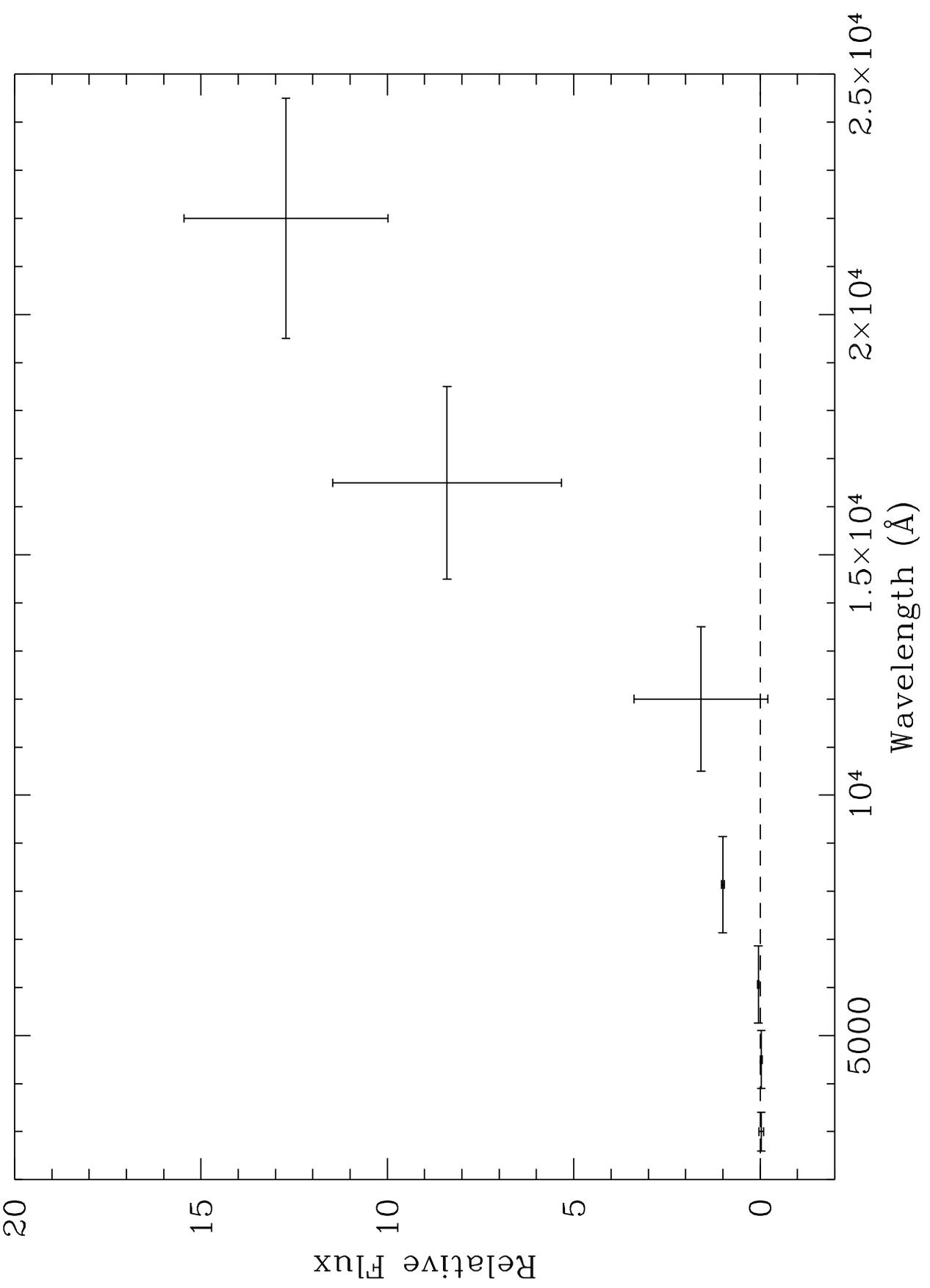
\end{document}